# Design and Analysis of Nonvolatile GSST-based Modulator Utilizing Engineered Mach-Zehnder Structure with Graphene Heaters

Sohrab Mohammadi-Pouyan, Shahram Bahadori-Haghighi, and, Afrooz Rafatmah

*Abstract*—photonic integrated circuits (PICs) are the foundation of on-chip optical technologies. Mach-Zehnder modulators (MZMs) are appealing building blocks of PICs which mostly rely on weak and volatile optical effects in materials. In contrast, phase change materials (PCMs) such as $Ge_2Sb_2Se_4Te_1$ (GSST) are promising candidates to realize efficient and nonvolatile reconfigurable optical devices. However, the phase transitions of PCMs are accompanied by large changes in the imaginary parts of their refractive indices which make the design of MZMs challenging. In this paper, two interesting design methods named as "loss-balancing" and "pre-equalization" are introduced to propose high-performance GSST-based MZMs. In this regard, a GSST-based waveguide with graphene as a microheater is proposed which plays the role of configurable active waveguide in both the introduced methods. According to the presented analysis, a nonvolatile MZM with active length of 4.725 μm and insertion loss of less than 2 dB at the wavelength of 1550 nm is realizable. Finally, the thermal simulation of the proposed GSST-based waveguide is performed so that the required voltages for amorphization (erasing) and crystallization (writing) processes are estimated to be 12 V and 4.3 V, respectively.

## I. INTRODUCTION

Photonic integrated circuits (PICs) are the leading platforms for on-chip integration of photonic systems in emerging applications such as neuromorphic computing and quantum information [1]. Reconfigurable components such as modulators and switches are key building blocks of PICs which convert the electrical signals into the optical ones. Low energy consumption, high bandwidth, small footprint, low insertion loss (IL) and complementary metal-oxide-semiconductor (CMOS) compatibility are essential characteristics of integrated modulators that have already been explored by researchers [2], [3]. However, most of the previously studied modulators and switches are based on weak thermo-optic, electro-optic and all-optical effects which lead to high energy consumption and large footprint. Besides, the aforementioned effects are all volatile so that a constant level of external power is required to retain the state of the modulator or switch. As a result, the energy consumption increases that is not desirable [4], [5].

Phase-change materials (PCMs) are promising candidates to realize efficient and nonvolatile reconfigurable optical devices. PCMs such as $VO_2$ [6], $Ge_2Sb_2Te_5$ (GST) have received growing attention due to their outstanding characteristics [7],[8]. GST exhibits a large complex refractive index change between its crystalline and amorphous phases so that strong modulation of light can be achieved within a small footprint. Furthermore, phase transitions of GST can take place by applying ultra-short optical or electrical pulses with low energy levels down to femtojoules [9]. In addition, the phase change of GST is nonvolatile and no power is required to sustain the new state. Another important advantage of PCMs such as GST is that they can be deposited on substrates without any concern about lattice mismatch issue [10]. The aforementioned characteristics are promising for various optical devices such as modulators/switches [11], [12] and optical memories [13], [14].

Despite attractive features, the associated loss of GST due to its small bandgap is not desirable which is specially critical when they are applied in cascaded PCM-based optical networks [15],[16], [17]. Substitution of some Te atoms by Se ones in GST results in an emerging PCM of $Ge_2Sb_2Se_4Te_1$ (GSST) with increased optical bandgap and reduced loss within the telecommunication bands around the wavelengths of 1310 nm and 1550 nm [15], [18]. The large refractive index contrast between the amorphous and crystalline states of GSST and its low optical loss in amorphous state, which is over 200 times smaller than that of GST [15], make it a promising material for optical devices such as Mach-Zehnder modulators (MZMs) with very small footprints and high modulation depth [19].

One of the PCM switching or modulation mechanisms that has recently been of interest is the electro-thermal method which enables on-chip integration [20]. Different kinds of materials have already been applied as the heater in this method. Metal heaters are appropriate choices for free-space reflective devices [20]. However, they introduce optical loss in the order of hundreds of dB/mm to the waveguide-based devices [20], [21]. In contrast, doped silicon heater is an ideal choice for PCM integration with the silicon-on-insulator (SOI) platforms but it cannot be easily applied to non-silicon waveguide structures [20]. On the other side, transparent conductive oxide (TCO) heaters are useful choices for visible spectrum but cannot be employed in the infrared range as they result in high optical loss [20], [22]. Graphene is a key solution for resolving such issues [20], [23].

Graphene is a two-dimensional (2D) array of carbon atoms arranged in a hexagonal lattice which exhibits intriguing electrical, thermal and optical properties [24]. The high thermal conductivity and high melting temperature of 4510 K make graphene a suitable candidate for heating purposes [25]. In other words, excellent heat transfer of graphene as a microheater leads to realization of high speed optical modulators [26]. Additionally, the optical loss of graphene within the infrared range can be reduced through Pauli-blocking effect. It has been reported that graphene microheaters for PCM switching can exhibit figure of merits higher than those of silicon or TCO heaters [20].

In this paper, we propose and design two types of high-performance GSST-based MZMs operating at the communication wavelength of 1550 nm. The modulator is made of a SOI waveguide with a GSST layer as the active material and a sheet of graphene as the microheater. In Section 2, the structure of the proposed waveguide is introduced and analyzed. The introduced waveguide is exploited to design the proposed MZMs in Section 3. Two interesting design methods of loss-balancing and pre-equalization are proposed in this section. According to the presented results, the proposed methods could pave the way for realization of low-loss GSST-based MZMs that are promising for nonvolatile optical applications. Finally, the electro-thermal simulation results of the GSST-based structure are presented in Section 4.

## II. PROPOSED ACTIVE WAVEGUIDE

### A. Proposed Waveguide

The perspective and cross-section view of the proposed waveguide are depicted in Fig. 1. As it is shown, the waveguide consists of a thin GSST layer deposited on a SOI ridge waveguide. A single layer graphene sheet is placed on the GSST layer as a heater to provide the required heat for GSST phase transition. Two Au/Ti pads are used at the two sides of the waveguide core to apply the external voltage for 1 heating process. There exists a low contact resistance between the Ti layer and graphene sheet. In addition, the electrodes are considered to be far away from the waveguide core to prevent possible excess optical loss.

When an external voltage is applied to the electrodes, an electrical current flows through the graphene sheet which causes heating of GSST that is originated by the electrical power dissipation. By applying an appropriate electrical voltage, the GSST temperature passes a critical point of 523 K where GSST changes from a low-loss amorphous phase to a lossy crystalline state that is called the writing process. The GSST layer will retain its new phase even if the supplied voltage is removed [27]. Inversely, when a bias voltage is applied to the waveguide with crystalline GSST over a period of time, the GSST temperature reaches its melting point where it returns to the amorphous phase that is known as the erasing process. Consequently, the refractive index of GSST is significantly modified during the phase transitions. Such refractive index variations are used to realize an efficient MZM. The ON-state of the proposed modulator is defined as the case with crystalline GSST while the OFF state is attributed to the amorphous GSST. It should also be noted that due to the non-volatile nature of the device, changes in the refractive index of graphene only affect the behavior of the modulator during the phase transition process while the voltage is supplied. Therefore, it is actually a transient factor that does not affect the steady states of the modulator.

### B. Optical Properties of the Proposed Waveguide

The width and height of the Si waveguide are respectively taken as $w = 480$ nm and $h = 220$ nm. The calculated real parts of the guided mode effective indices at the wavelength of 1550 nm in the ON- and OFF-states versus the GSST thickness are shown in Fig. 2 (a) and Fig. 2 (e), respectively. The corresponding electric field profiles of the waveguide for the TE, TM, and hybrid modes in the ON- and OFF-states are shown in Figs. 2 (b-d) and Figs. 2 (f-h), respectively. In the simulations, the complex dielectric constant of the monolayer graphene is calculated from Kubo formula [28].
The refractive indices of the GSST in the amorphous and crystalline phases are respectively $3.3258 + j1.8 \times 10^{-4}$ and $5.0830 + j0.350$ at the wavelength of 1550 nm [5].

According to the results depicted in Fig. 2, the modulator just supports fundamental TE and TM modes for $t_{GSST} < 40$ nm in both GSST phases. The access waveguide with TE propagating mode is weakly coupled to the TM-polarized guided mode of MZM while almost all the input power is coupled to the TE mode of modulator. However, by increasing the thickness of GSST, a hybrid mode will appear so that a notable portion of the propagating TE light of the access waveguide is coupled into this hybrid mode which causes excess loss. Therefore, the thickness of GSST must be lower than 40 nm.

## III. DESIGN AND CHARACTERIZATION OF THE MZMS

As it was mentioned in the previous section, the large refractive index change of GSST can lead to compact MZMs that are desirable for PICs. However, the variation in the imaginary part of the GSST refractive index during its phase transition is an obstacle for realization of high-performance MZMs. In other words, when a MZM with identical GSST-based arms is in the OFF state, there is no phase difference between propagating lights within the arms and constructive interference occurs at the output of MZM. However, when the modulator is set to ON state, GSST in the active arm changes to its crystalline phase which results in a phase shift and loss in the propagating light. Therefore, although the required phase shift between the two arms can be obtained at the output, the light amplitudes are not equal which prevents complete destructive interference at this state. In order to overcome such a problem, two MZM implementation techniques, which are denoted as 'loss-balancing' and 'pre-equalization' methods, are introduced and discussed in the following subsections.

*A. Loss-Balancing Method*

The 3D view of the proposed loss-balancing MZM is illustrated in Fig. 3, where two arms with different waveguides are used. the arm without any applied external voltage excitation is known as the passive arm which consists of a gold layer on the SOI waveguide. The active arm is exactly the waveguide proposed in Section 2. Also, $L_\pi$ represents the active length of the MZM arms.

In this method, the objective is to design geometrical parameters of the passive and active waveguides so that output light amplitudes become equal with a π-phase difference which ensures destructive interference in the ON state. In this regard, it is assumed that the two arms are of the same length $L_\pi$. Therefore, we first try to achieve the same imaginary parts of the effective indices for the waveguides of the two arms. As a result, the loss of the passive and active arms is balanced which ensures equal light amplitudes at their corresponding outputs in the ON state of the MZM. Then, $L_\pi$ is determined to achieve the required π phase difference at this state.

The optical properties of the waveguides highly depend on the thicknesses of GSST and Au layers. The calculated effective indices of the guided modes in the passive waveguide as functions of the Au thickness is shown in Fig. 4 (a). As it can be seen, the waveguide supports three plasmonic modes and only one TE mode which their corresponding electric field profiles are respectively depicted in Figs. 4 (b-e). The proposed modulator works based on the low-loss propagating TE mode. As the coupling from the TE mode of the access waveguide to the plasmonic TM modes of the passive arm is very weak, there is no concern about excitation of plasmonic modes. The effective index of the TE mode is designed to meet the mentioned criterion in the loss-balancing method.

The real ($n_{eff}$) and imaginary ($k_{eff}$) parts of the effective indices of the two waveguides for different thicknesses of the crystalline GSST and Au layers are respectively plotted in Fig. 5 (a) and Fig. 5 (b). In addition, the imaginary parts of the effective indices are separately depicted in Fig. 5 (c) for better clarity. As it is shown, the same imaginary parts or the same loss occurs at $t_{Au}$~65nm and $t_{GSST}$~7.5nm. In this case, the effective indices of the active waveguide in the amorphous and crystalline states of GSST are $n_{eff,Am} = 2.412+j3.2\times10^{-4}$ and $n_{eff,Cr} = 2.457+j0.001155$, respectively. The effective index of the passive waveguide is also calculated to be $n_{eff,Au} = 2.008+j0.001155$. Therefore, according to the following relation [29], the active arm length of the MZM, $L_\pi$, is calculated to be 1.726 μm at the wavelength of λ = 1550 nm.

$$L_\pi = \frac{\lambda}{2Re(n_{eff.cr} - n_{eff.Au})} \quad (1)$$

The optical performance of the designed MZM is now investigated. The transmission spectra of the MZM in the ON and OFF states are shown in Fig. 6 (a) with solid and dot-dashed lines, respectively. The ON- and OFF-state optical power distribution of the device at the wavelength of 1550 nm are also depicted in Fig. 6 (b) and Fig. 6 (c), respectively. As it is seen in Fig. 6 (a), in the ON-state, the output power transmission within the entire optical C-band is lower than -30 dB that is the consequence of loss-balancing method. However, the IL of the device that is the loss in the OFF state, is higher than 14 dB (denoted as uncompensated case in Fig.6 (a)) within the C-band which is the main drawback of this method. This is attributed to a significant phase difference, $\Delta\theta$, between the two arms at the OFF state which is calculated to be 161.86° according to the following equation:

$$\Delta\theta = 2\pi L_\pi \frac{Re(n_{eff.Am} - n_{eff.Au})}{\lambda} \quad (2)$$

The ON- and OFF-state output power transmission of the MZM for various Au and GSST thicknesses are respectively illustrated in Fig. 7 (a) and Fig. 7 (b). The regions where the effective modulation is achieved (i.e. $T_{ON}$ < -40 dB, $T_{OFF}$ > -20 dB) are indicated by a dashed curve. It is obvious that for effective modulation, the GSST layer thickness must be less than 15 nm. In order to clarify the aforementioned drawback of loss-balancing method, the phase difference between the amorphous and crystalline states of the active arm versus the GSST thickness is shown in Fig. 7 (c). It can be seen that for $t_{GSST}$ < 15nm (shaded region), the active arm experiences a maximum phase difference of almost 35° between the crystalline and amorphous GSST states. It implies that regardless of the GSST phase or MZM state, there is always a large phase difference between the two arms of the MZM which increases the IL.

In order to cope with this problem and achieve the better performance of the modulator, the loss-balancing MZM is modified to compensate $\Delta\theta$. In the following, the design details of the compensated loss-balancing modulator are discussed.

The proposed compensated loss-balancing MZM that is shown in Fig. 8, consists of two main parts including arm cores and an active phase compensator. The arm core section is used to provide the π-phase shift at the ON state and is exactly the same as the proposed loss-balancing MZM structure in Fig. 3. The phase compensator section is used to compensate the mentioned phase difference ($\Delta\theta$). The two arms of the phase compensator section are based on the GSST waveguide proposed in Section 2.

.

When the modulator is at the ON state, the arm cores result in the required π phase shift. Therefore, no additional phase difference should be induced by the active phase compensator and its two arms should be erased (the GSST layers in the phase compensator section should be in the amorphous state). On the other hand, when the modulator is at the OFF state, the arm of the compensator which is along with the passive waveguide (Au waveguide) must be turned on to compensate Δθ but the other waveguide remains off. The length of the phase compensator region ($L_\theta$) is calculated to be 15.48 µm at the wavelength of 1550 nm based on the following relation:

$$L_\theta = \frac{\lambda \Delta\theta}{2\pi Re(n_{eff.Cr} - n_{eff.Am})} \tag{3}$$

The transmission spectrum of the compensated loss-balancing MZM is also shown in Fig. 6 (a) with a dashed line. As it is shown, the IL varies between 6 dB and 9 dB for the entire optical C-band, which is reduced considerably compared to the IL of the uncompensated device. However, it should be noted that the improvement in the optical performance of the compensated MZM is at the cost of increased active length and energy consumption. The associated drawbacks of MZM in this method is mainly originated from the inhomogeneity of constituent materials of the two arms.

### B. Pre-equalization Method

The proposed homogeneous MZM for the pre-equalization method is shown in Fig. 9. The structure consists of a three-waveguide tapered directional coupler at the input which is able to inject the input light into the active and passive arms unequally. By selecting an appropriate power split ratio, the loss of the active arm in the crystalline phase can be compensated. In other words, it is possible to equalize light amplitudes at the output of the two arms which leads to an effective destructive interference at the ON state. The space gap between the waveguides of the input coupler determines the power split ratio [30]. Although unequal distribution of input power between the two arms is maintained at the OFF state with the amorphous GSST, the light amplitudes are added together by constructive interference at the output. Therefore, the IL originates from the intrinsic loss of materials, power splitter and bend losses not any phase difference.

Unlike the loss-balancing method where the layer thicknesses were determined to balance the loss of the two arms, the GSST thickness in the pre-equalization method can be selected to achieve a large effective index change of the waveguide during its phase transition. Therefore, according to Fig. 2, GSST thickness is assumed to be 30 nm which also ensures single mode operation. In this case, the guided mode effective indices in the crystalline and amorphous phases of GSST are respectively calculated to be $n_{eff,Cr}$ = 2.62581+j 0.0483468 and $n_{eff,Am}$ = 2.46451+j 1.022×10$^{-5}$.

The injected optical power into the passive arm ($T_P$) and active arm ($T_A$) of the MZM is related to the geometrical parameters of $g_P$, $g_A$ and $L_c$ in Fig. 9. Assuming $L_c$ = 5 µm and $g_A$ = 50 nm, the parameter $g_P$ is designed for optimal operation of MZM. The calculated transmitted power from the input to each arm of the MZM as a function of $g_P$ is plotted in Fig. 10 (a). In addition, the phase difference between the two arms caused by the power splitter ($\Delta\phi = \phi_A - \phi_P$) is also shown in Fig. 10 (a).

The design procedure is based on the transfer matrix method. The total optical transmission of the device is calculated as follows [31]:

$$T = \left|\frac{E_{out}}{E_{in}}\right|^2 \tag{4}$$

where, $E_{in}$ and $E_{out}$ are respectively the input and output optical electrical fields of the device. The output optical electric field is obtained from the following relation:

$$E_{out} = Y_{out}.M_{MZM}.Y_{in}.E_{in} \tag{5}$$

where, $Y_{in}$, $M_{MZM}$, and $Y_{out}$ are respectively the transfer matrices of the input power splitter, Mach-Zehnder arms, and output power combiner which are defined as:

$$Y_{in} = \begin{bmatrix} e^{-j\phi_P}\sqrt{T_P} \\ e^{-j\phi_A}\sqrt{T_A} \end{bmatrix} \tag{6}$$

$$Y_{out} = \begin{bmatrix} \frac{1}{\sqrt{2}} & \frac{1}{\sqrt{2}} \end{bmatrix} \tag{7}$$

$$M_{MZM} = \begin{bmatrix} e^{-j\phi_P}e^{-\frac{L_\pi}{2}\alpha_P} & 0 \\ 0 & e^{-j\phi_A}e^{-\frac{L_\pi}{2}\alpha_A} \end{bmatrix} \tag{8}$$

where $\alpha_A$ and $\alpha_P$ are the absorption coefficients of the active and passive arms, respectively which are calculated as $4\pi k_{eff}/\lambda$. Therefore, according to (4-8), the total transmission of the MZM can be calculated as:

$$T = \frac{1}{2}T_A e^{-L_\pi \alpha_A} + \frac{1}{2}T_P e^{-L_\pi \alpha_P} + \sqrt{T_A T_P}\, e^{-\left(\frac{\alpha_A + \alpha_P}{2}\right)L_\pi} \cos\left(\Delta\phi + 2\pi L_\pi \frac{\Delta n_{eff}}{\lambda}\right). \tag{9}$$

It should be reminded that $\Delta\phi$ is a constant phase difference between the two arms that is caused by the power splitter which can be considered as a portion of the required $\pi$ phase shift at the ON state to reduce the arm lengths. Therefore, the following relation should be held to satisfy the destructive interference condition in the ON state of the MZM.

$$\Delta\phi + 2\pi L_\pi \frac{\Delta n_{eff}}{\lambda} = \pi \tag{10}$$

Using (9) and (10), $L_\pi$ and the total transmission of the device in the ON-state ($T_{ON}$) as functions of $g_P$ are calculated and illustrated in Fig. 10 (b). It is obvious that the destructive interference condition is established at $g_P = 125$ nm with the corresponding $L_\pi$ of 4.2 μm. In this case, the additional phase difference caused by the power splitter is almost 30 degrees. However, it is important to note that such a phase difference also exists in the OFF state which increases IL. The phase difference due to the power splitter can be compensated if the passive arm is taken to be 50 nm longer than active arm. However, the required $L_\pi$ also increases to 4.725 μm.

The ON- and OFF-state optical transmission spectra and the corresponding optical power distribution of the propagating light at $\lambda = 1550$ nm are shown in Fig. 11 (a), Fig. 11 (b) and Fig. 11 (c), respectively. As it can be seen, the insertion loss of the MZM is lower than 2 dB and the total transmission in the ON-state is below -30 dB within the C-band, which demonstrate the better performance over the loss-balancing MZM that is attributed to the applied homogeneous arms in the pre-equalization method.

## IV. THERMAL SIMULATION OF THE DEVICE

In this section, phase transition of GSST actuated by heating through the graphene sheet is studied. In this context, multiphysics electro-thermal simulations have been performed based on material parameters including material density, thermal conductivity, and heat capacity. The thermal conductivities of GSST in the amorphous and crystalline phases are respectively assumed to be 0.17±0.02 W/mK and 0.43±0.04 W/mK and the heat capacity in the corresponding states is 1.45±0.05 MJ/m³K and 1.85±0.05 MJ/m³K, respectively [19]. Also, the density of GSST in amorphous and crystalline phases is respectively taken as 5900 kg/m³ and 6300 kg/m³ [32]. Moreover, the density, thermal conductivity, and heat capacity of $SiO_2$ are 2203 kg/m³, 1.38 W/mK and 746 J/kgK, respectively, The corresponding parameters for Si are respectively 2329 kg/m³, 130 W/mK and 700 J/kgK [33]. Graphene as the conductive material (with the electrical conductivity of 2.98 $10^6$ S.m$^{-1}$ [21]) for heating process is reported to have the density, thermal conductivity and heat capacity of 2267 kg/m³, 3000 W/mK and 2082 J/kgK, respectively [26].

Crystalline to amorphous phase transition or in other words the erasing process of GSST can take place by applying an appropriate pulse voltage with sufficient amplitude and duration to the graphene heater so that the GSST temperature reaches its melting point 900 K (amorphization temperature). The temperature of the coolest point of GSST during the erasing process is shown in Fig. 12 (a). As it can be seen, the phase transition from crystalline to amorphous state can be obtained by a pulse voltage of 12 V. The pulse duration required for the amorphization process is considered to be the time for raising the temperature of the GSST layer above its melting point. According to Fig. 12 (a), the amorphization time for loss-balancing and pre-equalization methods is simulated to be 75 ns and 480 ns, respectively. It is worth mentioning that the larger amorphization period for the pre-equalization method is due to the thicker GSST layer in its related structure. The 3D spatial temperature distribution in the proposed GSST waveguide designed with pre-equalization method and at the end of the amorphization process is shown in Fig. 12 (b). As it can be seen, the temperature of the whole GSST layer is higher than 900 K which ensures the crystalline to amorphous phase transition. Note that the temperature of the graphene region that is in contact with GSST is lower than that of the suspended graphene regions which confirms the fast heat transfer from graphene to the GSST layer.

The writing process is more challenging. The transition of GSST from amorphous to crystalline state occurs when its temperature is maintained above the critical point (~523 K) but below the melting point over a specific period of time. In this process, a voltage pulse train is required to keep the GSST temperature within the desired range [5]. The temperature of GSST during the crystallization process in the loss-balancing and pre-equalization methods are respectively shown in Fig. 12 (c) and Fig. 12 (d). The corresponding 1 MHz pulse trains with duty cycles of 50% are also shown by red lines. According to Fig. 12 (c) and Fig. 12 (d), the pulse amplitudes of 1.2 V and 4.3 V for loss-balancing and pre-equalization methods are respectively required to maintain the temperature of the GSST layer around 600 K. Note that as the melting temperature of graphene is 4510 K [25], neither in the amorphization nor the crystallization processes, the graphene layer is damaged. Once again, even by removing the applied voltage, the GSST layer sustains its crystalline phase which keeps the MZM in the ON state. Such a non-volatility, besides other advantages, makes our proposed modulator promising for a wide variety of optical applications in PICs.

## V. CONCLUSION

In summary, two new methods of loss-balancing and pre-equalization were introduced to design GSST-based MZMs. The optical modulation was performed by thermally triggered phase transition of GSST provided by the graphene heaters. The total active lengths of the GSST-based waveguides are calculated to be 17.2 µm and 4.725 µm for loss-balancing and pre-equalization methods, respectively. Although the IL of the compensated loss-balanced MZM is more than 6 dB within the C band, the pre-equalized modulator exhibits a low IL of less than 2 dB. The thermal simulation of the proposed GSST-based waveguide in the last section revealed that the amorphization (erasing) and crystallization (writing) processes can take place by external voltage pulses of 12 V and 4.3 V, respectively. The nonvolatile nature of such transitions makes our proposed MZMs promising for various applications in PICs.

FIGURES

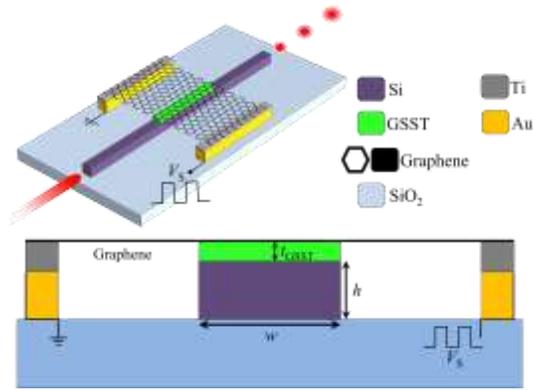

Fig. 1. The perspective and cross-section view of the proposed waveguide.

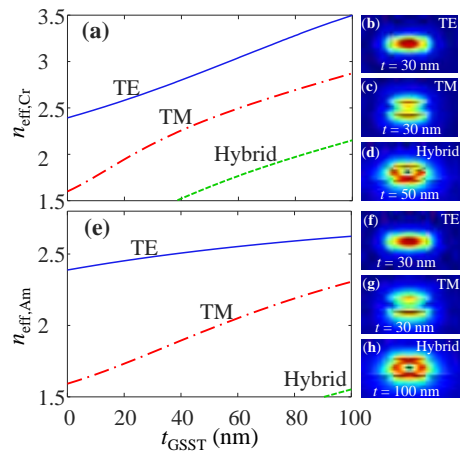

Fig. 2. Real parts of the TE, TM and hybrid mode effective indices of the proposed waveguide as a functions of $t_{GSST}$ in the (a) crystalline and (b) amorphous GSST states. The electric field profiles of the (c) TE, (d) TM and (e) hybrid modes when GSST is in the crystalline phase. The electric field profiles of the (f) TE, (g) TM and (h) hybrid modes when GSST is in the amorphous phase.

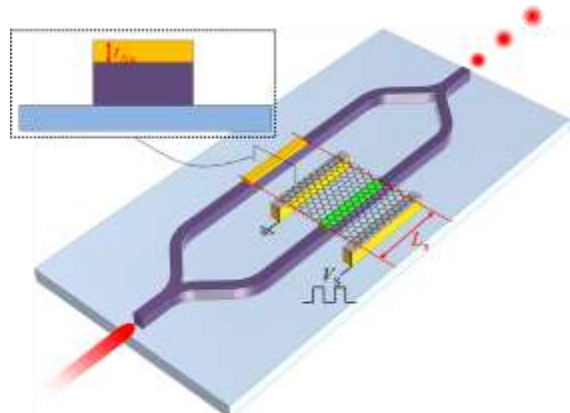

Fig. 3. The proposed MZM structure designed based on the loss-balancing method.

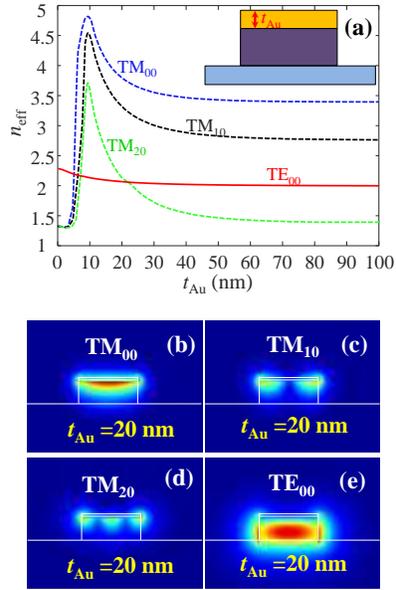

Fig. 4. (a) Effective indices of the passive waveguide modes as functions of the Au thickness. Electric field profiles of (b) $TM_{00}$, (c) $TM_{10}$, (d) $TM_{20}$ and (e) $TE_{00}$ modes.

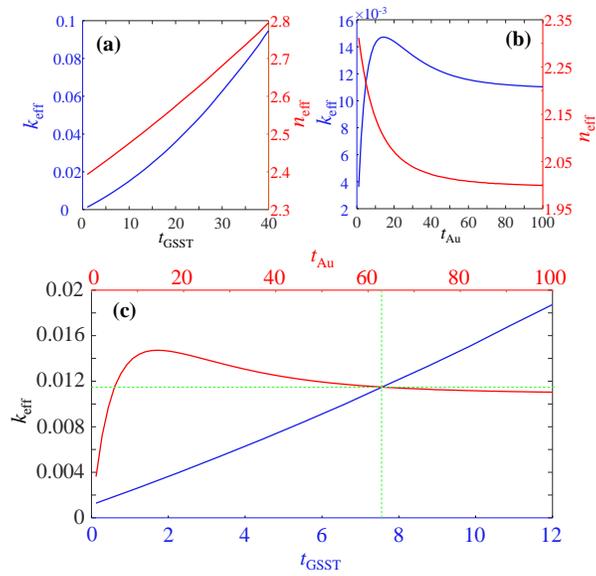

Fig. 5. Real and imaginary parts of guided mode effective indices of the (a) active and (b) passive waveguides versus crystalline GSST and Au thicknesses, respectively. (c) The imaginary parts of the effective indices versus GSST and Au thicknesses in the same graph.

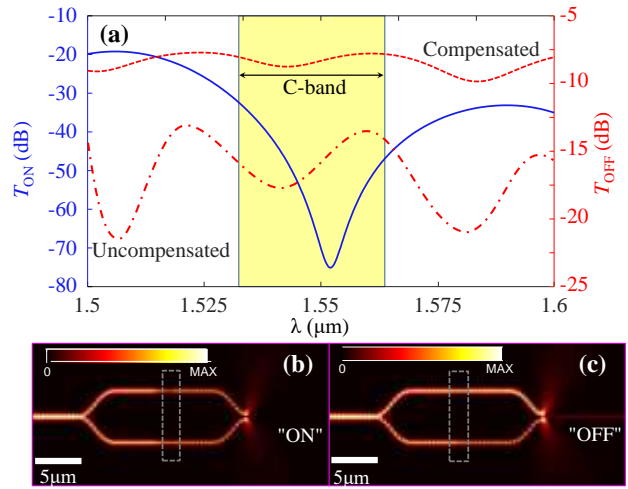

Fig. 6. (a) ON- and OFF-state output power transmission spectra of the uncompensated MZM and OFF-state transmission spectrum of the compensated MZM. Optical power distribution within the device in the (b) ON and (c) OFF states at the wavelength of 1550 nm.

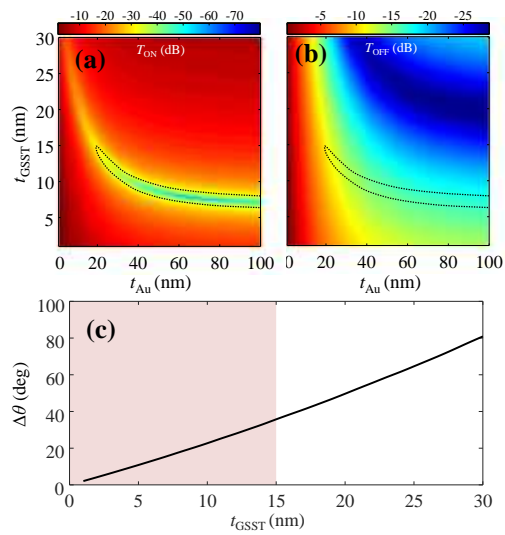

Fig. 7. (a) ON- and (b) OFF-state output power transmission profiles of the uncompensated MZM versus $t_{Au}$ and $t_{GSST}$. (c) Phase difference between the amorphous and crystalline states of the active arm as a function of the GSST thickness.

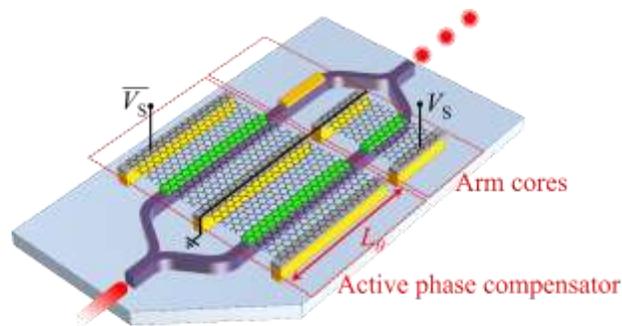

Fig. 8. 3D schematic of the proposed compensated loss-balancing MZM structure.

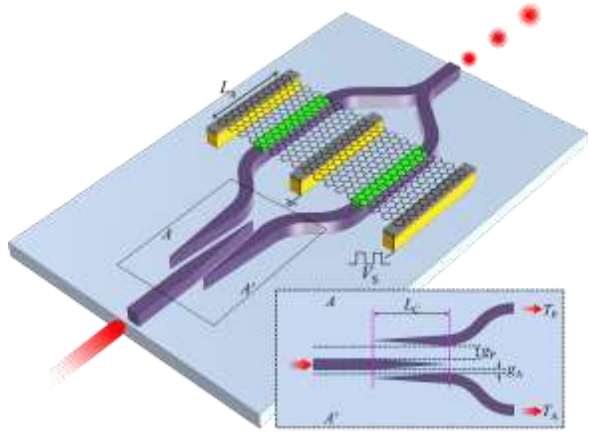

Fig. 9. 3D schematic of the proposed homogeneous MZM structure based on the pre-equalization method.

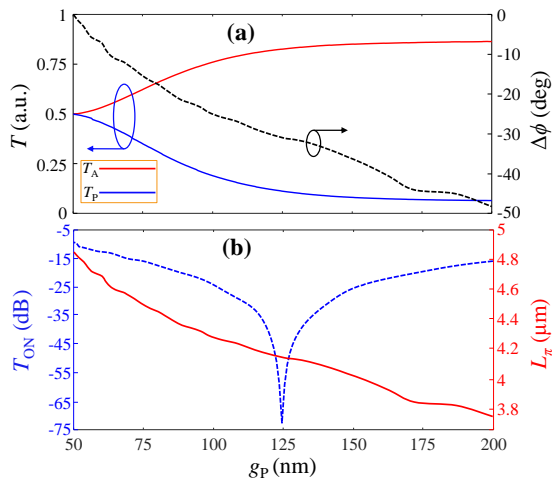

Fig. 10. (a) Injected optical power into the passive arm ($T_P$) and active arm ($T_A$) and the phase difference between the two arms caused by the power splitter as functions of $g_P$. (b) $L_\pi$ and ON-state transmission of the pre-equalized MZM versus $g_P$.

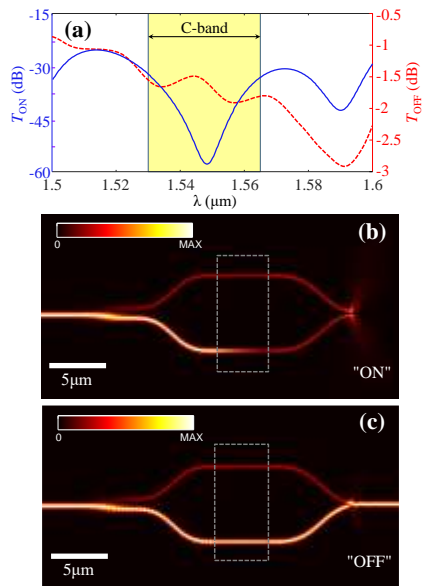

Fig. 11. (a) Output optical power spectra of the MZM designed with pre-equalization method in the ON and OFF states. The optical power distribution at the wavelength of 1550 nm in the (b) ON and (c) OFF states.

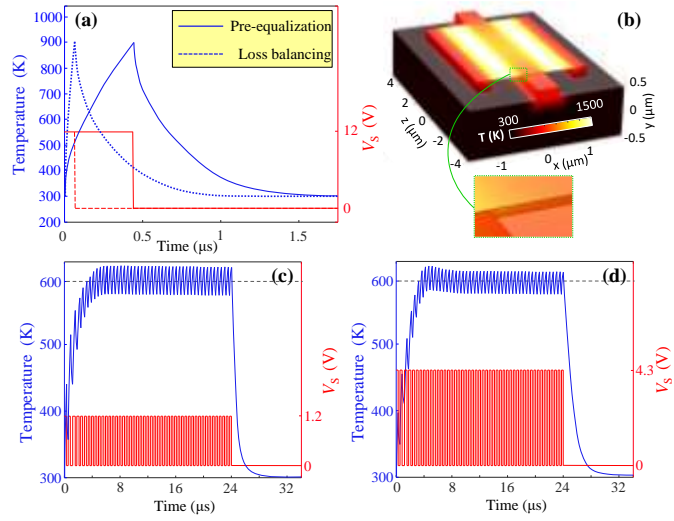

Fig. 12. (a) The applied external voltage signal for the amorphization process and the temperature of the GSST layer as functions of time. (b) The 3D spatial temperature distribution in the proposed waveguide of the MZM designed with the pre-equalization method. The temporal GSST temperature and applied external voltages for the crystallization process in the (c) loss-balancing and (d) pre-equalization methods.